\pdfoutput=1

\documentclass{article}

\def\arxivbuild{}

\usepackage[T1]{fontenc}

\usepackage{microtype}
\usepackage{graphicx}
\usepackage{subfigure}
\usepackage{booktabs}
\usepackage{multirow}
\usepackage{xcolor}
\usepackage{xr-hyper}
\usepackage{hyperref}
\usepackage{xurl}

\ifdefined\arxivbuild
  \usepackage[accepted]{mlsys2025}
  \makeatletter
  \renewcommand{\mlsys@appearing}{Preprint.}
  \makeatother
  \hypersetup{pdfsubject={}}
\else\ifdefined\sharestamp
  \usepackage[accepted]{mlsys2025}
  \usepackage{eso-pic}
  \AddToShipoutPictureFG{\AtPageLowerLeft{\makebox[\paperwidth]{\raisebox{0.25in}{\tiny\color{gray}Draft of
    \sharestamp. Do not distribute.}}}}
  \AddToShipoutPicture{\icmlruleroffset=\textheight\advance\icmlruleroffset by 5.2pt\color[rgb]{.7,.7,.7}%
    \AtTextUpperLeft{\put(\LenToUnit{-35pt},\LenToUnit{-\icmlruleroffset}){\icmlruler{\icmlrulercount}}}}
\else
  \usepackage{mlsys2025}
\fi\fi

\usepackage{amsmath}
\usepackage{amssymb}
\usepackage{xspace}
\usepackage{array}
\newcolumntype{L}[1]{>{\raggedright\arraybackslash}p{#1}}
\usepackage{enumitem}
\usepackage{tikz}
\usetikzlibrary{arrows.meta,positioning,fit,calc,patterns}

\usepackage{pgfplots}
\pgfplotsset{compat=1.18}
\usepgfplotslibrary{groupplots,fillbetween}
\usepackage{sansmath}

\definecolor{cSentinel}{HTML}{2A78D6}
\definecolor{cFence}{HTML}{EB6834}
\definecolor{cDeepEp}{HTML}{6B6A66}
\definecolor{cMegaMoE}{HTML}{3D3C39}
\definecolor{cDeepEpGemm}{HTML}{898781}
\definecolor{cTwoSided}{HTML}{6B6A66}
\definecolor{cNcclEp}{HTML}{8F8E89}
\definecolor{cBf16}{HTML}{6B6A66}
\definecolor{cMxfp8}{HTML}{6250D6}
\definecolor{cNvfp4}{HTML}{008300}
\definecolor{cFp8}{HTML}{E34948}
\definecolor{cInk}{HTML}{0B0B0B}
\definecolor{cInk2}{HTML}{52514E}
\definecolor{cGrid}{HTML}{E4E3DF}
\definecolor{cAxis}{HTML}{9C9B96}
\definecolor{cEp8}{HTML}{86B6EF}
\definecolor{cEp16}{HTML}{3987E5}
\definecolor{cEp32}{HTML}{1C5CAB}
\definecolor{cEp64}{HTML}{0D366B}
\definecolor{cXfer}{HTML}{1BAF7A}
\definecolor{cCopy}{HTML}{EDA100}
\definecolor{cSync}{HTML}{6250D6}
\definecolor{cRoute}{HTML}{C9C8C3}

\newcommand{\plotfont}{\fontsize{7}{8.4}\fontfamily{qhv}\selectfont\sansmath}
\newcommand{\plottitlefont}{\fontsize{8}{9.6}\fontfamily{qhv}\selectfont\sansmath}

\pgfplotsset{
  paper/.style={
    width=\linewidth, height=4.2cm,
    font=\plotfont,
    label style={font=\plotfont, text=cInk2},
    tick label style={font=\plotfont, text=cInk2},
    title style={font=\plottitlefont, text=cInk, inner sep=0pt, yshift=-1pt, text height=1.6ex, text depth=0.4ex},
    axis line style={cAxis, line width=0.4pt},
    tick style={cAxis, line width=0.4pt},
    major tick length=2pt, minor tick length=0pt,
    axis x line*=bottom, axis y line*=left,
    ymajorgrids, grid style={cGrid, line width=0.35pt},
    every axis plot/.append style={line width=0.9pt, mark size=1.6pt, line join=round},
    legend style={draw=none, fill=none, font=\plotfont, text=cInk, inner sep=0pt,
      /tikz/every even column/.append style={column sep=9pt}},
    legend image code/.code={
      \draw[mark repeat=2, mark phase=2] plot coordinates {(0cm,0cm) (0.3cm,0cm) (0.6cm,0cm)};},
    legend cell align=left,
    log basis x=2,
    /pgf/number format/1000 sep={},
  },
  tokens axis/.style={xmode=log, xlabel={tokens per rank}, xtick={1,4,16,64,256,1024},
    xticklabels={1,4,16,64,256,1K}},
  speedup ticks/.style={ytick={1,2,3}, yticklabels={1\texttimes,2\texttimes,3\texttimes},
    extra y ticks={1}, extra y tick labels={}, extra y tick style={major tick length=0pt, ymajorgrids,
      major grid style={cDeepEp, line width=0.6pt}}},
  sSentinel/.style={color=cSentinel, solid, mark=*, mark size=1.7pt,
    mark options={solid, fill=cSentinel, draw=white, line width=0.4pt}},
  sFence/.style={color=cFence, solid, mark=square*, mark size=1.5pt,
    mark options={solid, fill=cFence, draw=white, line width=0.4pt}},
  sDeepEp/.style={color=cDeepEp, densely dashed, line width=0.75pt, mark=triangle*, mark size=1.9pt,
    mark options={solid, fill=cDeepEp, draw=white, line width=0.4pt}},
  sMegaMoE/.style={color=cMegaMoE, densely dashed, line width=0.75pt, mark=diamond*, mark size=1.9pt,
    mark options={solid, fill=cMegaMoE, draw=white, line width=0.4pt}},
  sDeepEpGemm/.style={color=cDeepEpGemm, densely dashed, line width=0.75pt, mark=triangle*, mark size=1.9pt,
    mark options={solid, rotate=180, fill=cDeepEpGemm, draw=white, line width=0.4pt}},
  sTwoSided/.style={color=cTwoSided, densely dashed, mark=triangle*, mark options={solid, fill=cTwoSided}},
  sNcclEp/.style={color=cNcclEp, dashed, mark=diamond*, mark options={solid, fill=cNcclEp}},
}
\tikzset{end label/.style={font=\plotfont, inner sep=1pt}}

\newcommand{\sys}{\texttt{NVLinkOneSided}\xspace}
\newcommand{\fence}{\textsc{Fence}\xspace}
\newcommand{\sent}{\textsc{Sentinel}\xspace}

\newcommand{\deepep}{DeepEP\xspace}

\newcommand{\dedg}{DE+DG}
\newcommand{\loadimb}{Imba.}

\newcommand{\nranks}{P}          %
\newcommand{\topk}{k}            %
\newcommand{\ntok}{T}            %
\newcommand{\tstep}{\hat{\ntok}}%
\newcommand{\nexp}{E}            %
\newcommand{\eloc}{E_{\ell}}     %
\newcommand{\us}{\,\textmu s\xspace}

\newcommand{\tok}[1]{\textsf{T#1}}
\newcommand{\rk}[1]{\textsf{R#1}}
\newcommand{\ex}[1]{\textsf{E#1}}

\ifdefined\arxivbuild
  
  \newcommand{\Supplementary}{}
\else
  
  \newcommand{\Supplementary}{Supplementary }
\fi

\newcommand{\prelim}[1]{#1}
\newcommand{\figplaceholder}[2]{%
  \fbox{\parbox[c][#1][c]{0.95\linewidth}{\centering\small\color{gray}#2}}}

\newcommand{\drawio}[3]{\IfFileExists{figures/drawio/#1.pdf}%
  {\includegraphics[width=#2]{figures/drawio/#1.pdf}}{\figplaceholder{#3}{\detokenize{#1}}}}

\setlist{leftmargin=*,itemsep=1pt,topsep=2pt}

\newcommand{\algpanel}[3]{\refstepcounter{algorithm}\label{#2}%
  \hrule height.8pt depth0pt\kern2pt
  \noindent{\normalsize\textbf{Algorithm~\thealgorithm}~#1}\par
  \kern2pt\hrule\kern2pt
  #3\par
  \kern2pt\hrule height.8pt depth0pt}

\newenvironment{algnotation}{\begin{description}[font=\normalfont,leftmargin=1em,labelsep=0.3em,itemsep=0pt,
  parsep=0pt,topsep=1pt,partopsep=0pt]}{\end{description}}

\renewcommand{\algorithmiccomment}[1]{\hfill$\triangleright$~#1}

\makeatletter
\let\mlsys@makecaption\@makecaption
\long\def\@makecaption#1#2{\mlsys@makecaption{#1}{#2}\def\@tempa{table}\ifx\@captype\@tempa\vskip 4pt\fi}
\makeatother

\mlsystitlerunning{One-Sided Expert-Parallel All-to-All for Mixture-of-Experts Inference over NVLink}

\newcommand{\papertitle}{One-Sided Expert-Parallel All-to-All for Mixture-of-Experts Inference over NVLink}

\newcommand{\paperauthors}{%
\begin{mlsysauthorlist}
\mlsysauthor{Bo Li}{nv}
\mlsysauthor{Chulian Zhang}{nv}
\mlsysauthor{Or Biri}{nv}
\mlsysauthor{Anthony Chang}{nv}
\mlsysauthor{Anton Korzh}{nv}
\mlsysauthor{Noam Duvdevani}{nv}
\mlsysauthor{XingFei Xi}{nv}
\mlsysauthor{Daniel Stokes}{nv}
\mlsysauthor{Zhenhuan Chen}{nv}
\mlsysauthor{Ludwig Schneider}{nv}
\mlsysauthor{Artem Y. Polyakov}{nv}
\mlsysauthor{Julien Demouth}{nv}
\mlsysauthor{Chandler Zhou}{nv}
\end{mlsysauthorlist}
\mlsysaffiliation{nv}{NVIDIA}
\mlsyscorrespondingauthor{Bo Li}{lbo@nvidia.com}
\mlsyscorrespondingauthor{Chulian Zhang}{chulianz@nvidia.com}
\mlsyskeywords{Mixture of Experts, Expert Parallelism, All-to-All, NVLink, Symmetric Memory, LLM Inference}}

\begin{document}

\twocolumn[
\mlsystitle{\papertitle}

\paperauthors

\vskip 0.3in

\begin{abstract}
Expert parallelism (EP) shards the experts of a Mixture-of-Experts (MoE) layer across GPUs, so every MoE layer
requires two all-to-all communications, namely dispatch and combine. Compared with training, inference processes relatively
few tokens per GPU at a time, and at small batch sizes fixed costs that
do not grow with the number of tokens, rather than data movement, dominate the latency. We present \sys, a one-sided EP all-to-all for
GPUs within a single NVLink domain.
Each source rank pushes its tokens, without first exchanging metadata, into rank-major buffers on the target GPUs,
and the grouped GEMMs read them in place instead of from a per-expert copy. We provide two protocols for the
synchronization that the communication requires, namely \fence and \sent, for high-throughput and low-latency scenarios
respectively. On Blackwell GPUs, \sys completes dispatch and combine \prelim{1.2--3.0$\times$} faster than \deepep,
with the largest speedups at small batch sizes. The resulting MoE layer is up to \prelim{2.4$\times$} faster than
\deepep with DeepGEMM. It even outperforms the fused DeepGEMM MegaMoE at up to \prelim{128} tokens per GPU, making it
still competitive for decoding. We expose \sys through a modular interface, so that it is agnostic to the activation
quantization format, decoupled from the MoE compute backend and easy to integrate into inference frameworks.

\end{abstract}
]

\printAffiliationsAndNotice{}

\section{Introduction}
\label{sec:intro}

Mixture-of-Experts (MoE) models activate only a few of their many experts for each
token~\cite{shazeer2017outrageously,lepikhin2021gshard,fedus2022switch}. They are commonly served with data-parallel (DP)
attention and expert-parallel (EP) MoE layers, a deployment strategy popularized by DeepSeek-V3~\cite{deepseekv3}. In the attention layers, every rank serves its
own requests and processes only their tokens. In the MoE layers, every rank holds a subset of the experts, so every
MoE layer contains two all-to-all communications.
\emph{Dispatch} sends each token to the ranks that own its experts, and \emph{combine} returns the expert outputs to
the source rank of the token.

Compared with training, inference processes few tokens per GPU at a time. Prefill is usually split into chunks of
thousands to tens of thousands of tokens~\cite{agrawal2024sarathi,deepinfer2026}, and decoding must meet a service-level objective on the time per
output token, which bounds its batch to a few hundred tokens per rank. DeepSeek's DeepInfer~\cite{deepinfer2026}, for
example, decodes DeepSeek-V4.1-Flash at about 250 tokens per second per user while serving 64 requests per
rank; each request adds three draft tokens, so a step processes 256 tokens per rank.

At large batches, dispatch and combine are bound by the communication bandwidth, so the goal is to use as much of it
as possible and to eliminate unnecessary data movement. At small batches, however, every
microsecond matters~\cite{shen2026microsecond}. At a few tokens per rank, the bytes a rank sends would take less than a microsecond at
the theoretical NVLink bandwidth, yet measured round trips of dispatch and combine take tens of microseconds (\S\ref{sec:eval}) in every
MoE layer: the time goes to fixed costs that do not grow with the number of tokens, such as synchronization between
producers and consumers on different GPUs. In common designs such as NCCL~\cite{nccl} and \deepep~\cite{deepep}, the sender writes data into an
intermediate buffer, from which the receiver copies or regroups it into the layout that the next kernel needs; this
extra pass over the data adds latency at small batches and consumes memory bandwidth at large ones.

Within one NVLink domain, symmetric memory lets a rank load from and store to a peer's memory directly, so data
can be placed exactly where its consumer reads it. We present \sys, a one-sided EP all-to-all built on this direct
access. The routing
decisions of a rank alone determine where each of its tokens goes, so dispatch can push the tokens into the memory of
the receivers without first sending routing metadata such as token counts or expert ids. The dispatch \emph{workspace}
of each rank, a buffer in symmetric memory, is divided into one region per source rank, so that different ranks do not interfere when writing to the same target rank;
within its region, a source rank places each token in one slot, however many experts of the target rank the token
selects. We call this layout \emph{rank-major}. The grouped GEMMs of the MoE layer read the tokens of each expert
directly from this workspace through slot indices, without materializing the tokens into per-expert order.
Dispatch and combine still need synchronization between producers and consumers, for which we provide two protocols
targeting high-throughput and low-latency scenarios respectively. \fence publishes per-peer flags after a memory fence. \sent uses the data
itself as the flag. We expose \sys through the general EP communication interface of FlashInfer,
\texttt{MoEEpCommunication}, which makes it agnostic to the
activation quantization format and easy to integrate with different inference frameworks and MoE backends.

The rest of the paper is organized as follows. Section~\ref{sec:design} presents the one-sided communication and the
rank-major layout that MoE compute reads. Section~\ref{sec:sync} explains the synchronization protocols and
implementation details. Section~\ref{sec:eval} evaluates \sys and the MoE layer built on it on Blackwell GPUs, analyzes
the two protocols and compares both with existing implementations.

\section{Background}
\label{sec:background}

\subsection{Mixture-of-Experts and expert parallelism}
\label{sec:bg-moe}

\begin{figure}[t]
  \centering
  \drawio{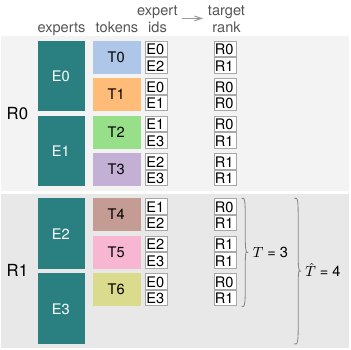}{2.336in}{2.33in}
  \caption{Example used throughout the paper: two ranks, experts \ex{0}, \ex{1} on \rk{0} and \ex{2}, \ex{3} on
  \rk{1}, and top-2 routing. \rk{0} holds tokens \tok{0}--\tok{3} and \rk{1} holds \tok{4}--\tok{6}. Beside each token are
  its two expert ids and, through the routing (arrow), the target rank of each: both experts of \tok{1} are on \rk{0}, both
  experts of \tok{3} and \tok{5} are on \rk{1}, and every other token reaches both ranks. The braces on \rk{1} mark
  its number of tokens, \mbox{$\ntok = 3$}, and the largest number of tokens of any rank, \mbox{$\tstep = 4$}
  (Table~\ref{tab:notation}).}
  \label{fig:routing}
\end{figure}

An MoE layer replaces the dense feed-forward network with $\nexp$ experts and a router~\cite{shazeer2017outrageously,
lepikhin2021gshard,fedus2022switch}, which selects $\topk$ experts for each token and weights their outputs. Under
EP the experts are split over $\nranks$ GPUs (\emph{ranks}), $\eloc = \nexp/\nranks$ per rank, while
attention runs in DP, so each rank holds its own $\ntok$ tokens. The layer then takes three steps.
\emph{Dispatch} sends each token to the ranks that own its experts, each rank runs its local experts on the tokens it
received, and \emph{combine} returns the weighted expert outputs to the source rank of each token, which sums them. Dispatch
is a multicast whose destinations depend on the router's choices, and combine is the reverse, a gather with a reduction.
Every rank may send tokens to every other rank, and the amounts depend on the routing, so both are all-to-all
exchanges with variable sizes. Figure~\ref{fig:routing} shows the example used throughout the paper.

\subsection{NVLink domains and symmetric memory}
\label{sec:bg-nvlink}

A Blackwell GPU has 18 fifth-generation NVLink links of 100\,GB/s each, 900\,GB/s per direction or 1.8\,TB/s in
total. NVLink switches fully connect the eight GPUs of an HGX B300 board, and a GB200 or GB300 NVL72 rack forms one
72-GPU NVLink domain. Beyond the domain, each GPU communicates through one RDMA NIC, a ConnectX-8 of 800\,Gb/s, that
is, 100\,GB/s per direction, roughly an order of magnitude less~\cite{nv-blackwell-ultra-blog}.

Within the domain, GPUs address each other's memory directly. With CUDA's Virtual Memory Management, each rank
allocates physical memory, exports a handle, and maps every peer's allocation into its own address space; the result
is a \emph{symmetric memory} region in which a kernel accesses peer memory with ordinary loads, stores and
atomics~\cite{nccl227blog,pytorch-symmem}. RDMA offers remote access through a different interface: the initiator, a CPU thread or, with
GPU-initiated networking, a GPU thread~\cite{nv-ibgda-blog}, posts work requests to the NIC, names registered
regions by key, and is notified of completion through completion queues or a write with immediate data~\cite{ibta-spec};
transfers are efficient for messages of kilobytes. Table~\ref{tab:nvlink-rdma} lists the differences that matter for an
all-to-all of small, data-dependent messages. \sys moves data into and out of peer memory with
16-byte stores and loads from many threads and detects completion through the GPU memory model
(\S\ref{sec:sync}).

\begin{table}[t]
  \centering
  \scriptsize
  \caption{Remote access within an NVLink domain and over RDMA (Blackwell, per GPU).}
  \label{tab:nvlink-rdma}
  \setlength{\tabcolsep}{4pt}
  \begin{tabular}{@{}lll@{}}
    \toprule
    & NVLink domain & RDMA NIC \\
    \midrule
    Initiator & any thread, load/store & work request to the NIC \\
    Transfer size & store ($\le$32\,B) or TMA copy & message, KB scale \\
    Ordering & PTX fences, system scope & per queue pair \\
    Arrival in a message & whole 32-bit words & unspecified \\
    \bottomrule
  \end{tabular}
\end{table}

A remote store does not wait for the peer: the thread moves on at once, and the data reach the peer later, possibly
out of order with other stores. GPU memory is weakly ordered~\cite{lustig2019ptx}: a consumer on another GPU that sees a producer's write is
guaranteed to see the producer's earlier writes only if the producer ordered them with a system-scope release and the
consumer reads with acquire semantics. In addition, each aligned 32-bit word stored to peer memory is observed whole
or not at all~\cite{ptx-isa}. NCCL's LL protocol relies on a stronger guarantee: an 8-byte store of 4 bytes of data
and a 4-byte flag arrives whole~\cite{nccl,demystifying-nccl}. RDMA offers no
counterpart: the order of placement within an RDMA Write is undefined, and the responder may rely on the data only
after a later completion, such as a write with immediate data~\cite{ibta-spec,rfc5040}.

\section{One-Sided Communication with Rank-Major Layout}
\label{sec:design}

This section describes one-sided communication, the rank-major layout and how MoE compute reads it.
Table~\ref{tab:notation} summarizes the notation.

\begin{table}[t]
  \centering
  \scriptsize
  \caption{Notation and the corresponding parameter names in the implementation.}
  \label{tab:notation}
  \setlength{\tabcolsep}{4pt}
  \begin{tabular}{@{}lll@{}}
    \toprule
    Symbol & Meaning & Name in the implementation \\
    \midrule
    $\nranks$ & EP ranks & \texttt{ep\_size} \\
    $\topk$ & experts per token & \texttt{top\_k} \\
    $\nexp$ & experts of the layer & \texttt{num\_experts} \\
    $\eloc$ & experts per rank & \texttt{num\_local\_experts} \\
    $\ntok$ & tokens of a rank in the step & \texttt{local\_num\_tokens} \\
    $\tstep$ & largest $\ntok$ across ranks & \texttt{runtime\_max\_tokens\_per\_rank} \\
    \bottomrule
  \end{tabular}
\end{table}

\subsection{One-sided communication}
\label{sec:design-onesided}

\begin{figure*}[t]
  \centering
  \drawio{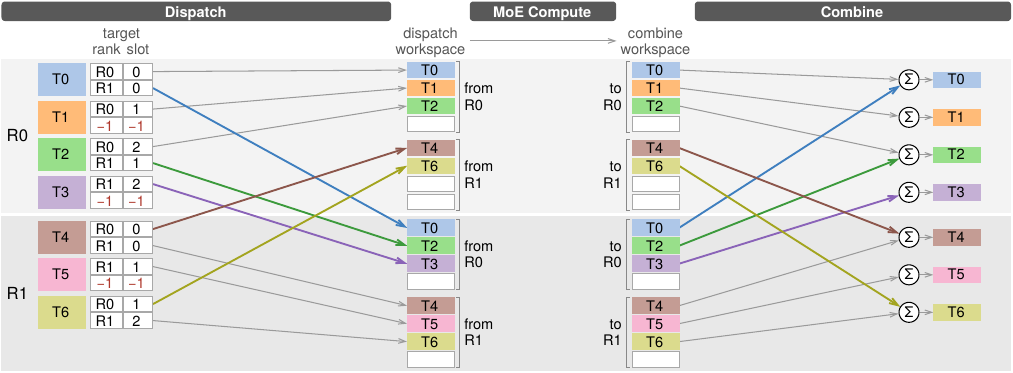}{\textwidth}{2.48in}
  \caption{Data path for the routing of Figure~\ref{fig:routing}, with the \fence synchronization protocol. Dispatch:
  each source rank stores its tokens into its own region of the dispatch workspaces of the target ranks and records
  the target rank and slot of each selection; \tok{1} selects \ex{0} and \ex{1}, both on \rk{0}, so it takes one slot
  there and its second selection records no slot ($-1$). MoE compute turns the tokens from each source rank into
  partial results at the same slots of the combine workspace. Combine: each source rank loads the partial results from
  the target ranks and slots recorded in dispatch and adds them. White slots are empty.}
  \label{fig:datapath}
\end{figure*}

Each rank routes its own tokens, so it knows locally which target ranks each token must reach. In dispatch, the source
rank \emph{pushes} each token directly, at a slot that it chooses itself, into the dispatch workspace of every target
rank, which lies in symmetric memory, and records the target rank and slot of each selection
(Figure~\ref{fig:datapath}). No metadata, such as token counts or expert ids, travels ahead of the data. MoE
compute leaves the partial results in a second workspace, the combine workspace, from which combine returns them.

\subsection{Rank-major layout}
\label{sec:design-layout}

Source ranks push into a target rank concurrently, so their stores must not interfere. Slots drawn from one shared
pool would need atomics on a remote counter, or offsets precomputed from the counts of all source ranks, which would
require a metadata exchange. Instead, each dispatch workspace has one region per source rank, which only that source
rank writes: it takes the next slot by atomically incrementing its \emph{slot counter} for that target rank, and its
slots form a dense prefix of the region. We call this layout \emph{rank-major}. A region has $\tstep$ slots, which
every rank knows because inference engines with DP attention share the batch size information every step; the dispatch
workspace is sized at initialization for the maximum number of tokens a rank can process. A token goes to each target
rank once, however many experts of that rank it selects, with all $\topk$ expert ids and router weights in its slot.
Some grouped GEMM implementations, such as DeepGEMM~\cite{deepgemm}, read the tokens of each expert contiguously, so
the corresponding dispatch delivers an \emph{expert-major} layout instead, with one slot per selected
expert.\footnote{The low-latency kernels of \deepep V1 push each token once per selected expert into an expert-major
buffer. \deepep V2.5 pushes each token once per target rank into a rank-major communication buffer; with
\texttt{do\_expand}, which DeepGEMM requires, it then copies each token once per selected local expert into an
expert-major GEMM input.} The rank-major layout has two benefits over it:

\textbf{De-duplication.} A token is sent at most $\min(\topk, \nranks)$ times instead of $\topk$ times as in a
per-expert transfer; for example, with $\topk = 16$ on eight GPUs, it sends at most half the bytes, whatever the
routing. The saving is larger when $\nranks$ is small relative to $\topk$, or when a token chooses several experts
located on the same rank.

\textbf{Smaller workspaces.} The rank-major dispatch and combine workspaces each hold $\nranks\tstep$ slots. An
expert-major buffer must hold every selection that a rank may receive. If every local expert may receive every token,
as in the low-latency kernels of \deepep V1, it needs $\eloc$ times as many slots. Since a token selects at most
$\min(\topk, \eloc)$ experts of one rank, \deepep V2.5 still needs $\min(\topk, \eloc)$ times as many. For
DeepSeek-V4.1-Flash on eight GPUs with up to 4096 tokens per rank, the hidden states take 160\,MiB in our dispatch
workspace ($8 \times 4096$ slots of 5120 bytes in MXFP8) and 320\,MiB in our combine workspace (BF16), besides a few
MiB for scaling factors, expert ids and router weights. \deepep V2.5 with DeepGEMM needs a 242\,MiB communication
buffer, shared by dispatch and combine, and an expert-major GEMM input whose hidden states alone take 1020\,MiB, around
$\topk = 6$ times the 160\,MiB of our dispatch workspace (the excess is due to alignment padding). All these buffers
grow linearly with the EP size, so memory becomes more pressing at large EP.

\begin{figure*}[t]
  \centering
  \drawio{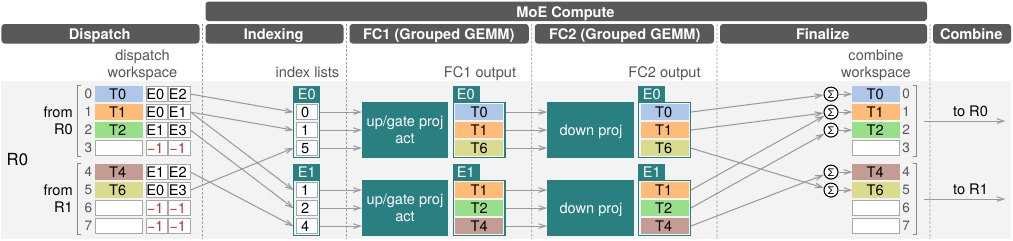}{\textwidth}{1.60in}
  \caption{The MoE layer on \rk{0}, detailing Figure~\ref{fig:datapath}; MoE compute comprises indexing, FC1, FC2 and
  finalize, and slots are numbered 0--7 across the two regions. The indexing kernel turns the expert ids of the
  received slots into per-expert lists of slot indices (empty slots 3, 6 and 7 carry invalid ids); FC1 gathers
  tokens through these lists and multiplies them by the expert weights, FC2 multiplies the FC1 output by its own
  weights, and finalize adds the outputs of each token into its slot of the combine workspace.}
  \label{fig:pipeline}
\end{figure*}

\subsection{MoE compute on the rank-major layout}
\label{sec:design-bridge}

An MoE layer on the rank-major layout runs dispatch, the MoE compute kernels (an indexing kernel, the grouped GEMMs FC1 and FC2, and
finalize) and combine (Figure~\ref{fig:pipeline}). The tokens stay at their slots of the dispatch workspace, and the
grouping by expert is expressed as slot indices.

\textbf{Indexing.} A small indexing kernel selects the ids of the local experts, skips the slots marked invalid
(details in \S\ref{sec:sync}), and builds per-expert lists of slot indices, plus the inverse map, from each slot to
the FC2 output rows of its token, for finalize. FC1 gathers its
input through these indices, so no separate pass permutes the received tokens before FC1.

\textbf{FC1 and FC2.} FC1 applies the up and gate projections and the activation of each expert to its tokens, and
FC2 applies the down projection of the expert to the result. FC1 loads its input by index directly from the
dispatch workspace: the Prims-TS grouped GEMM of FlashInfer~\cite{flashinfer-primsts} gathers it with TMA
\texttt{gather4} instructions, each of which loads 128-byte slices of four non-contiguous slots, and never loads empty
slots. FC1 writes its output grouped by expert, and FC2 reads it in that order.

\textbf{Finalize and combine.} Finalize gathers the FC2 outputs of each token through the inverse map, scales them
by the router weights, sums them over the local experts of the token, and writes the result into the combine workspace
that the communication layer exposes, at the slot of the token. The finalized partial results are thus back in the
rank-major layout of dispatch, so combine needs no metadata either: the source rank \emph{pulls} the partial results
of each token from the combine workspaces of its target ranks, at the slots recorded in dispatch, and adds them.

\section{Implementation and Synchronization Protocols}
\label{sec:sync}

\begin{figure}[t]
  \centering
  \drawio{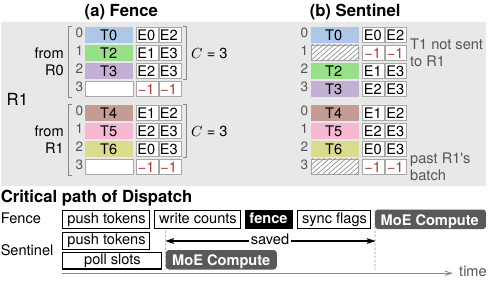}{\linewidth}{1.87in}
  \caption{The dispatch workspace of \rk{1} at the end of dispatch. (a) \fence gathers the tokens of each source rank
  into a prefix of its region, so each source rank sends its slot count $C$; the slots past the
  count get invalid expert ids ($-1$). (b) \sent pushes token $i$ of each source rank into slot $i$ of its region,
  so every slot signals its own arrival; a slot without a token keeps the sentinel payload (hatched) and
  gets invalid expert ids. Bottom: the critical path of dispatch until MoE compute starts; \sent shortens it, which pays off at small
  batches.}
  \label{fig:protocols}
\end{figure}

Kernels in one stream are ordered: a kernel sees every write to local memory made by the kernels before it. Remote memory accesses have no such guarantee: the stores of a source rank reach a peer while
the kernels of the peer run, in no particular order (\S\ref{sec:bg-nvlink}). One-sided dispatch and combine therefore
need two producer--consumer synchronizations to prevent races:
\begin{itemize}[itemsep=0pt,parsep=0pt]
  \item \textbf{End of dispatch.} Before a target rank starts MoE compute, every source rank must have finished
  storing its tokens into the dispatch workspace of the target rank.
  \item \textbf{Start of combine.} Before a source rank reads the partial results of its tokens, every target rank
  must have finished MoE compute and written the results into its combine workspace.
\end{itemize}
\sys provides two synchronization protocols: \fence for high-throughput and \sent for low-latency scenarios. They differ in how a source rank assigns slots and how a target rank learns that its slots
are complete (Figure~\ref{fig:protocols}).

\subsection{\fence synchronization}
\label{sec:sync-fence}

Each CUDA thread block handles the dispatch and the combine of one token, and the blocks of a rank run in parallel
(Algorithm~\ref{alg:fence}). A block takes the slot of its token in each target rank by atomically incrementing the
slot counter for that rank. When a block has sent its token, it increments another local counter of finished blocks,
and the last block to finish synchronizes for the whole rank, so the rank executes one system-scope fence per step.
This block publishes the slot counts to their target ranks, executes a system-scope release fence
(\texttt{fence.release.sys}), and writes a flag to every peer rank. It spins until the flags of all peer ranks have
arrived, which ends the dispatch kernel. A small kernel marks the expert ids past each published count invalid, so
that the slots past the prefix take no part in MoE compute. Combine synchronizes in the same way, except that the
first block alone writes the flags while every block spins on the flags of all peers and executes an acquire fence
(\texttt{fence.acquire.sys}) before it pulls the finalized partial results of its token from the target ranks and
slots recorded in dispatch.

\begin{figure*}[t]
  \small
  \begin{minipage}[t]{0.485\textwidth}
  \algpanel{\sys \fence protocol}{alg:fence}{%
  \begin{algnotation}
      \item[{Thread block:}] token $i$ on source rank $s$
      \item[{$W_d[r]$, $W_c[r]$:}] dispatch and combine workspaces of rank $r$
      \item[{$C[r][s]$, $F[r][s]$:}] slot count and flag of source rank $s$ on rank $r$
      \item[{$\mathit{cnt}[t]$:}] local slot counter for target rank $t$, initially 0
      \item[{$\mathit{done}$:}] local counter of finished blocks, initially 0
      \item[{$f$:}] flag value of the step
    \end{algnotation}
    \begin{algorithmic}[1]
      \STATE \textbf{Dispatch}
      \STATE $D \gets$ target ranks of token $i$
      \FORALL{$t \in D$}
        \STATE $j_t \gets \mathrm{atomicAdd}(\mathit{cnt}[t], 1)$ \COMMENT{pick slot on $t$}
        \STATE $W_d[t][s][j_t] \gets$ token $i$ \COMMENT{push token}
      \ENDFOR
      \IF{$\mathrm{atomicAdd}(\mathit{done}, 1) = \ntok - 1$}
        \STATE $C[r][s] \gets \mathit{cnt}[r]$ for all $r$ \COMMENT{write slot counts}
        \STATE \texttt{fence.release.sys}
        \STATE $F[r][s] \gets f$ for all $r$ \COMMENT{sync via completion flags}
        \STATE wait until all $F[s][r] = f$
      \ENDIF
      \FORALL[sanitize expert ids]{$r$, $j \ge C[s][r]$}
        \STATE $W_d[s][r][j].\mathit{expert\_ids} \gets -1$
      \ENDFOR
      \item[]
      \STATE \textbf{Combine}
      \IF{first block}
        \STATE $F[r][s] \gets f + 1$ for all $r$ \COMMENT{sync via completion flags}
      \ENDIF
      \STATE wait until all $F[s][r] = f + 1$
      \STATE \texttt{fence.acquire.sys}
      \STATE $p_t \gets W_c[t][s][j_t]$ for all $t \in D$ \COMMENT{pull tokens in parallel}
      \STATE output of token $i \gets \sum_{t \in D} p_t$ \COMMENT{reduction}
    \end{algorithmic}}
  \end{minipage}\hfill
  \begin{minipage}[t]{0.485\textwidth}
  \algpanel{\sys \sent protocol}{alg:sentinel}{%
  \begin{algnotation}
      \item[{Thread block:}] token slot $i < \tstep$ on source rank $s$
      \item[{$L[r]$, $I[r]$, $M[r]$:}] landing, expert-id and combine slots of rank $r$
      \item[{$W_d[r]$, $W_c[r]$:}] dispatch and combine workspaces of rank $r$
      \item[{$\sigma$:}] sentinel, initial value of $L$, $I$ and $M$
    \end{algnotation}
    \begin{algorithmic}[1]
      \STATE \textbf{Dispatch}
      \STATE $D \gets$ target ranks of token $i$, $\emptyset$ if $i \ge \ntok$
      \STATE $L[t][s][i] \gets$ token $i$ for all $t \in D$ \COMMENT{push token}
      \STATE $I[r][s][i] \gets$ expert ids of token $i$ ($-1$ if $r \notin D$) for all $r$
      \FORALL{source ranks $r$}
        \STATE wait until no word of $I[s][r][i]$ equals $\sigma$ \COMMENT{poll expert ids}
        \IF{$I[s][r][i] \ne -1$}
          \STATE wait until no word of $L[s][r][i]$ equals $\sigma$ \COMMENT{poll token}
          \STATE $W_d[s][r][i] \gets L[s][r][i]$; \; $L[s][r][i] \gets \sigma$ \COMMENT{copy, reset}
        \ENDIF
      \ENDFOR
      \item[]
      \STATE \textbf{Combine}
      \STATE $M[r][i][s] \gets W_c[s][r][i]$ for all $r$ with $I[s][r][i] \ne -1$ \COMMENT{push partial results}
      \STATE $I[s][r][i] \gets \sigma$ for all $r$ \COMMENT{reset expert-id slots}
      \FORALL[poll partial results]{$t \in D$}
        \STATE wait until no word of $M[s][i][t]$ equals $\sigma$
        \STATE $p_t \gets M[s][i][t]$; \; $M[s][i][t] \gets \sigma$ \COMMENT{load, reset}
      \ENDFOR
      \STATE output of token $i \gets \sum_{t \in D} p_t$ \COMMENT{reduction}
    \end{algorithmic}}
  \end{minipage}
\end{figure*}

\subsection{\sent synchronization}
\label{sec:sync-sentinel}

\fence ends every dispatch with a serial chain: atomic counting for the slot indices, a system-scope fence, and the
publishing of and spinning on completion flags. \sent shortens the critical path by letting the data signal their own
arrival. Source ranks push the tokens into landing slots, each 32-bit word of which is initialized with a sentinel
value, a bit pattern that the data do not take or that the producer avoids: BF16 payload words, for example, use
\texttt{0xFFFFFFFF}, two BF16 NaNs, and the producer stores such a word as two \texttt{0x7FFF} NaNs instead. A target
rank polls the words of a landing slot; since each word arrives whole (\S\ref{sec:bg-nvlink}), a slot whose words all
differ from the sentinel is complete without a fence. The target rank copies it into the same slot of its dispatch
workspace, which MoE compute reads as with \fence, and resets it to the sentinel. Unlike NCCL's LL
protocol~\cite{nccl}, which interleaves flag words with the data, \sent adds no flag words: the data themselves are the
flags.\footnote{Several code bases call this Lamport synchronization, after Lamport's happened-before
relation~\cite{lamport1978time}: the arrival of the data itself orders the producer's stores before the consumer's
loads.}

The routing decides which slots receive a token, so a target rank must tell a slot still in flight from one that will
never be written. Token $i$ of source rank $s$ therefore always lands in slot $i$ of region $s$, and for each of its
$\tstep$ slots every source rank writes expert ids to every rank, invalid ones ($-1$) where no token goes
(Figure~\ref{fig:protocols}b): every expert-id slot arrives and tells whether a payload follows. Expert-id slots use
the sentinel \texttt{0x88888888}, which differs from $-1$.

Combine pushes too: once the MoE compute of a target rank has finished, it stores the partial result of slot $i$ of
region $s$ into a combine slot on source rank $s$ without waiting for its peers, and the block of token $i$ polls these
slots like landing slots and adds the partial results.

The kernel that polls a slot consumes it and resets it to the sentinel. A rank writes a landing or combine slot on a
peer again only after receiving data that the peer sends in a later kernel, the partial result of the landing slot or
the expert ids of the next step, so one set of these slots suffices. Expert-id slots also reach ranks that send nothing
back, so they alternate between two sets: a rank finishes dispatch only after the expert ids of every rank have
arrived, so no rank starts a step before every rank has finished the step before last.

\sent's costs grow with the batch: each rank writes, polls and resets $\nranks\tstep$ expert-id slots per step, and every
received word is checked, copied and reset, which shrinks the advantage of \sent (\S\ref{sec:eval-sync}).

\subsection{Kernel implementation details}
\label{sec:impl}

\sys is implemented as a set of CUDA kernels behind FlashInfer's \texttt{MoEEpCommunication}
interface. Dispatch moves tokens after quantization: MoE compute follows dispatch, so the
quantized tokens naturally take part in the computation. Dispatch therefore supports every quantization
format of the MoE compute backends (e.g., BF16, FP8, MXFP8 and NVFP4) and sends the payloads of each token: the
quantized hidden states, their scaling factors, and the expert ids and router weights from the routing. MoE compute
outputs BF16, so combine usually transfers BF16. The kernels with the \fence protocol can also quantize the partial
results to FP8 before combine and dequantize them before the reduction, which saves combine bandwidth at a potential loss of
precision.

Both kernels keep as many bytes in flight over NVLink as they can, through two kinds of parallelism:
\begin{itemize}[itemsep=0pt,parsep=0pt]
  \item \textbf{Data-level parallelism (DLP).} Each thread uses the widest vectorized loads and stores that the
  payload alignment allows, 16 bytes.
  \item \textbf{Instruction-level parallelism (ILP).} In dispatch, a thread loads each vector from local memory once
  and stores it to all target ranks of the token in an unrolled loop, so the remote stores of the token are in flight
  together. In combine, a thread first loads the vectors of all target ranks of the token into registers and then adds
  them pairwise in a tree, which keeps the dependency chain short.
\end{itemize}
The size of the rank-major workspaces depends only on the number of tokens, so the whole MoE layer can be captured in
a CUDA graph, in which dispatch, the MoE compute kernels and combine use Programmatic Dependent Launch
(PDL)~\cite{cuda-pdl} to overlap each kernel with its predecessor. Every synchronization is a
point where a failed peer stalls the others, so every wait has a timeout and traps with an error message instead of
hanging the engine.

\section{Evaluation}
\label{sec:eval}

We measure the all-to-all alone (\S\ref{sec:eval-latency}), the whole MoE layer with it (\S\ref{sec:eval-moe}), and
both in a larger NVLink domain (\S\ref{sec:eval-scaling}).

\subsection{Setup}
\label{sec:eval-setup}

Sections~\ref{sec:eval-latency}--\ref{sec:eval-moe} run on eight NVLink-connected B300 GPUs and
\S\ref{sec:eval-scaling} on a GB200 NVL72 rack to evaluate large EP. We test the routed-expert layers of
DeepSeek-V4.1-Flash~\cite{dsv41flash} and Kimi-K3~\cite{kimik3} (Table~\ref{tab:models}) at 1 to 4096 tokens per rank,
with \emph{perfect} routing: the $\topk$ selections of each token go to consecutive ranks and spread evenly over their
experts.

\begin{table}[t]
  \centering
  \footnotesize
  \caption{Configuration of the MoE layers of the tested models, with the bytes per token and target rank.}
  \label{tab:models}
  \setlength{\tabcolsep}{4pt}
  \begin{tabular}{@{}lrr@{}}
    \toprule
    & DeepSeek-V4.1-Flash & Kimi-K3 \\
    \midrule
    Hidden size & 5120 & 3584\rlap{\textsuperscript{a}} \\
    Experts, top-$\topk$ & 384, 6 & 896, 16 \\
    Expert intermediate size & 2304 & 3072 \\
    Dispatch bytes (MXFP8) & 5328 & 3824 \\
    Combine bytes (BF16) & 10240 & 7168 \\
    \bottomrule
    \multicolumn{3}{@{}p{0.95\linewidth}@{}}{\textsuperscript{a}\,Latent MoE; the original hidden size is 7168.}
  \end{tabular}
\end{table}

Our MoE layer pairs \sys with the Prims-TS grouped GEMMs of \S\ref{sec:design-bridge}, with MXFP4 weights and MXFP8
activations; where a single number stands for \sys, it is that of the faster protocol. The baselines are \deepep
V2.5~\cite{deepep} with the settings of its own tests; DeepGEMM MegaMoE~\cite{deepgemm}, one persistent kernel that
overlaps NVLink transfers with computation; and \deepep + DeepGEMM (DE+DG), the unfused baseline of MegaMoE's test.
\Supplementary Appendix~\ref{app:method} describes the measurement.

\subsection{All-to-all performance}
\label{sec:eval-latency}

\begin{figure}[t]
  \centering
  \begin{tikzpicture}
\begin{groupplot}[
  group style={group size=2 by 2, horizontal sep=10pt, vertical sep=5pt, xticklabels at=edge bottom,
    yticklabels at=edge left, ylabels at=edge left},
  paper, tokens axis, xlabel={}, scale only axis, width=0.39\linewidth,
  xmin=0.7, xmax=5800, xtick={1,4,16,64,256,1024,4096}, xticklabels={1,4,16,64,256,1K,4K},
  unbounded coords=jump,
]
\nextgroupplot[title={DeepSeek-V4.1-Flash}, height=0.72in,
  ymode=log, ymin=14, ymax=800, ytick={20,50,100,200,500}, yticklabels={20,50,100,200,500},
  ylabel={round trip (\textmu s)},
  legend to name=a2alegend, legend columns=2, transpose legend]
  \addplot[sSentinel] table[x=T, y=rt] {data/series/a2a_v41flash_sentinel.dat};
  \addlegendentry{\sys\ \sent}
  \addplot[sFence] table[x=T, y=rt] {data/series/a2a_v41flash_fence.dat};
  \addlegendentry{\sys\ \fence}
  \addplot[sDeepEp] table[x=T, y=rt] {data/series/a2a_v41flash_deepep.dat};
  \addlegendentry{\deepep}
\nextgroupplot[title={Kimi-K3}, height=0.72in,
  ymode=log, ymin=14, ymax=800, ytick={20,50,100,200,500}]
  \addplot[sSentinel] table[x=T, y=rt] {data/series/a2a_kimik3_sentinel.dat};
  \addplot[sFence] table[x=T, y=rt] {data/series/a2a_kimik3_fence.dat};
  \addplot[sDeepEp] table[x=T, y=rt] {data/series/a2a_kimik3_deepep.dat};
\nextgroupplot[height=0.46in, ymin=0.7, ymax=3.3, speedup ticks, ylabel={speedup}]
  \addplot[sSentinel] table[x=T, y=speedup] {data/series/a2a_v41flash_sentinel.dat};
  \addplot[sFence] table[x=T, y=speedup] {data/series/a2a_v41flash_fence.dat};
\nextgroupplot[height=0.46in, ymin=0.7, ymax=3.3, speedup ticks, yticklabels={}]
  \addplot[sSentinel] table[x=T, y=speedup] {data/series/a2a_kimik3_sentinel.dat};
  \addplot[sFence] table[x=T, y=speedup] {data/series/a2a_kimik3_fence.dat};
\end{groupplot}
\node[anchor=south, inner sep=0pt] at ($(group c1r1.north)!0.5!(group c2r1.north)+(0,13pt)$)
  {\pgfplotslegendfromname{a2alegend}};
\node[anchor=north, inner sep=0pt, font=\plotfont, text=cInk2]
  at ($(group c1r2.south)!0.5!(group c2r2.south)+(0,-11pt)$) {tokens per rank};
\end{tikzpicture}
  \caption{\sys completes dispatch and combine faster than \deepep at every batch. Round trip on eight B300 GPUs;
  bottom: speedup over \deepep.}
  \label{fig:latency}
\end{figure}
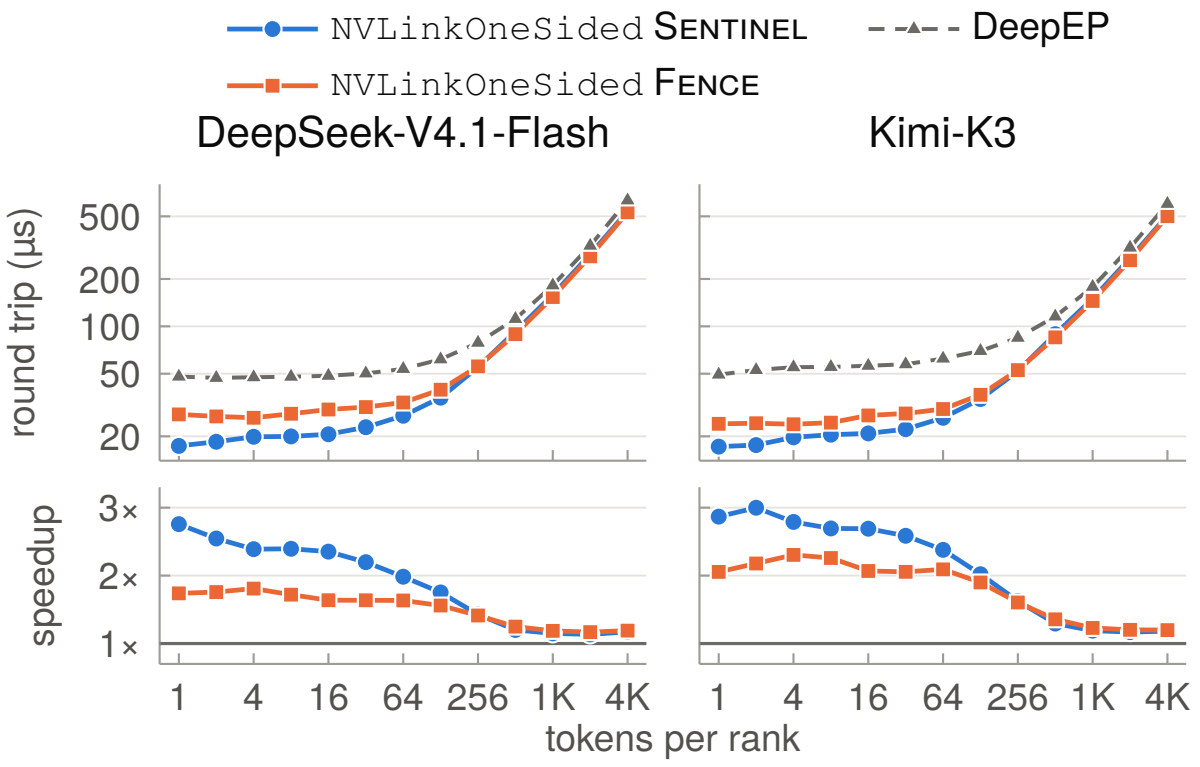

\sys is faster than \deepep at every batch, most where fixed costs dominate (Figure~\ref{fig:latency}): by
\prelim{1.75--3.00$\times$} up to 128 tokens per rank and by \prelim{1.17--1.63$\times$} from 256. At one token per
rank, the round trip on DeepSeek-V4.1-Flash takes \prelim{17.4}\us with \sent, \prelim{27.6}\us with \fence and
\prelim{47.9}\us with \deepep; at 4096 tokens, \fence reaches \prelim{68--71\%} of the 900\,GB/s unidirectional NVLink
bandwidth.

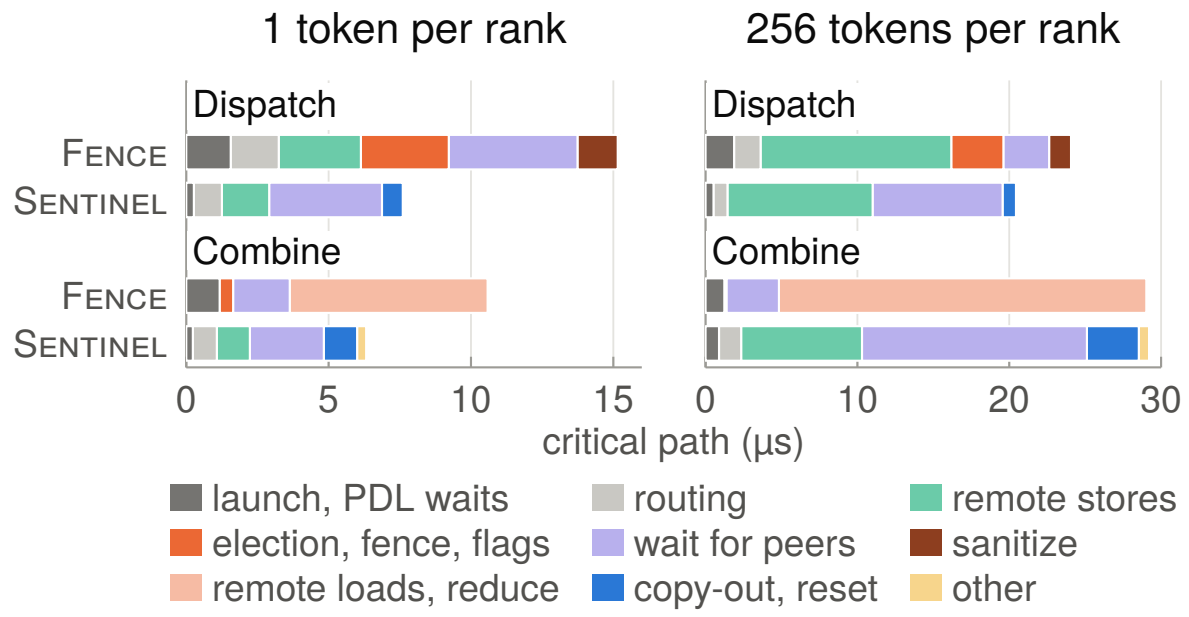
\begin{figure}[t]
  \centering
  \begin{tikzpicture}
\def\stepsrow{(mod(\coordindex,4) < 2 ? 40.5 - 9*mod(\coordindex,4) : 31.5 - 9*mod(\coordindex,4))}
\def\stepsbars{%
  \addplot[fill=cInk2!80] table[x=launch, y expr=\stepsrow] {data/series/steps_phase_v41flash.dat};
  \addplot[fill=cRoute] table[x=routing, y expr=\stepsrow] {data/series/steps_phase_v41flash.dat};
  \addplot[fill=cXfer!65] table[x=stores, y expr=\stepsrow] {data/series/steps_phase_v41flash.dat};
  \addplot[fill=cFence] table[x=fence, y expr=\stepsrow] {data/series/steps_phase_v41flash.dat};
  \addplot[fill=cSync!45] table[x=wait, y expr=\stepsrow] {data/series/steps_phase_v41flash.dat};
  \addplot[fill=cFence!60!black] table[x=sanitize, y expr=\stepsrow] {data/series/steps_phase_v41flash.dat};
  \addplot[fill=cFence!45] table[x=loads, y expr=\stepsrow] {data/series/steps_phase_v41flash.dat};
  \addplot[fill=cSentinel] table[x=copyout, y expr=\stepsrow] {data/series/steps_phase_v41flash.dat};
  \addplot[fill=cCopy!45] table[x=other, y expr=\stepsrow] {data/series/steps_phase_v41flash.dat};
  \node[anchor=west, font=\plotfont, text=cInk, fill=white, inner sep=1pt] at (axis cs:0,49) {Dispatch};
  \node[anchor=west, font=\plotfont, text=cInk, fill=white, inner sep=1pt] at (axis cs:0,22) {Combine};}
\begin{groupplot}[
  group style={group size=2 by 1, horizontal sep=12pt, yticklabels at=edge left},
  paper, scale only axis, width=0.365\linewidth, height=54pt, xmin=0,
  ymin=0, ymax=54, ytick={40.5,31.5,13.5,4.5}, yticklabels={\fence,\sent,\fence,\sent},
  ymajorgrids=false, xmajorgrids, major y tick style={draw=none},
  every axis plot/.append style={draw=white, line width=0.4pt},
]
\nextgroupplot[xbar stacked, bar width=6.5pt, title={1 token per rank}, xmax=16, xtick={0,5,10,15},
  every axis plot/.append style={skip coords between index={4}{8}}]
  \stepsbars
\nextgroupplot[xbar stacked, bar width=6.5pt, title={256 tokens per rank}, xmax=30, xtick={0,10,20,30},
  every axis plot/.append style={skip coords between index={0}{4}}]
  \stepsbars
\end{groupplot}
\node[anchor=north, font=\plotfont, text=cInk2, inner sep=0pt]
  at ($(group c1r1.south west)!0.5!(group c2r1.south east)+(0,-11pt)$) {critical path (\textmu s)};
\node[anchor=north, font=\plotfont, text=cInk2, inner sep=0pt]
    at ($(group c1r1.south west)!0.5!(group c2r1.south east)+(0,-21pt)$) {%
  \begin{tabular}{@{}l@{\hspace{6pt}}l@{\hspace{6pt}}l@{}}
    \textcolor{cInk2!80}{\rule{6pt}{5pt}}~launch, PDL waits & \textcolor{cRoute}{\rule{6pt}{5pt}}~routing &
      \textcolor{cXfer!65}{\rule{6pt}{5pt}}~remote stores \\
    \textcolor{cFence}{\rule{6pt}{5pt}}~election, fence, flags & \textcolor{cSync!45}{\rule{6pt}{5pt}}~wait for peers &
      \textcolor{cFence!60!black}{\rule{6pt}{5pt}}~sanitize \\
    \textcolor{cFence!45}{\rule{6pt}{5pt}}~remote loads, reduce & \textcolor{cSentinel}{\rule{6pt}{5pt}}~copy-out, reset &
      \textcolor{cCopy!45}{\rule{6pt}{5pt}}~other
  \end{tabular}};
\end{tikzpicture}
  \caption{\sent shortens the work around the wait for peers, not the wait. Critical path inside the dispatch and
  combine kernels, from IKET timestamps; DeepSeek-V4.1-Flash on eight B300 GPUs. Launch, PDL waits include \fence's
  counter-reset kernel; election, fence, flags: \fence's serial chain; sanitize: \fence's kernel that invalidates the
  expert ids past the slot counts, mostly overlapped with dispatch; copy-out, reset: \sent's handling of received
  slots; other: kernel exit and \sent's local reduction.}
  \label{fig:sync}
\end{figure}

\paragraph{\fence or \sent.}
\label{sec:eval-sync}
\sent is faster at decoding batches and \fence at large ones, with the crossover at 256 tokens per rank
(Figure~\ref{fig:latency}, bottom). In-Kernel Event Tracing (IKET)~\cite{cutlass-iket} shows why
(Figure~\ref{fig:sync}). At one token per rank, both protocols wait about equally long for their peers, and \sent
saves the work around the wait: its dispatch skips \fence's serial chain and extra kernels and takes half as long, and
its combine pushes each partial result when it is ready instead of waiting for all peers and pulling with remote loads.
As the batch grows, the transfers dominate with either protocol, while \sent's per-slot work, copying out and resetting
every received slot and running more combine blocks than fit on the GPU at once, cancels its saving.

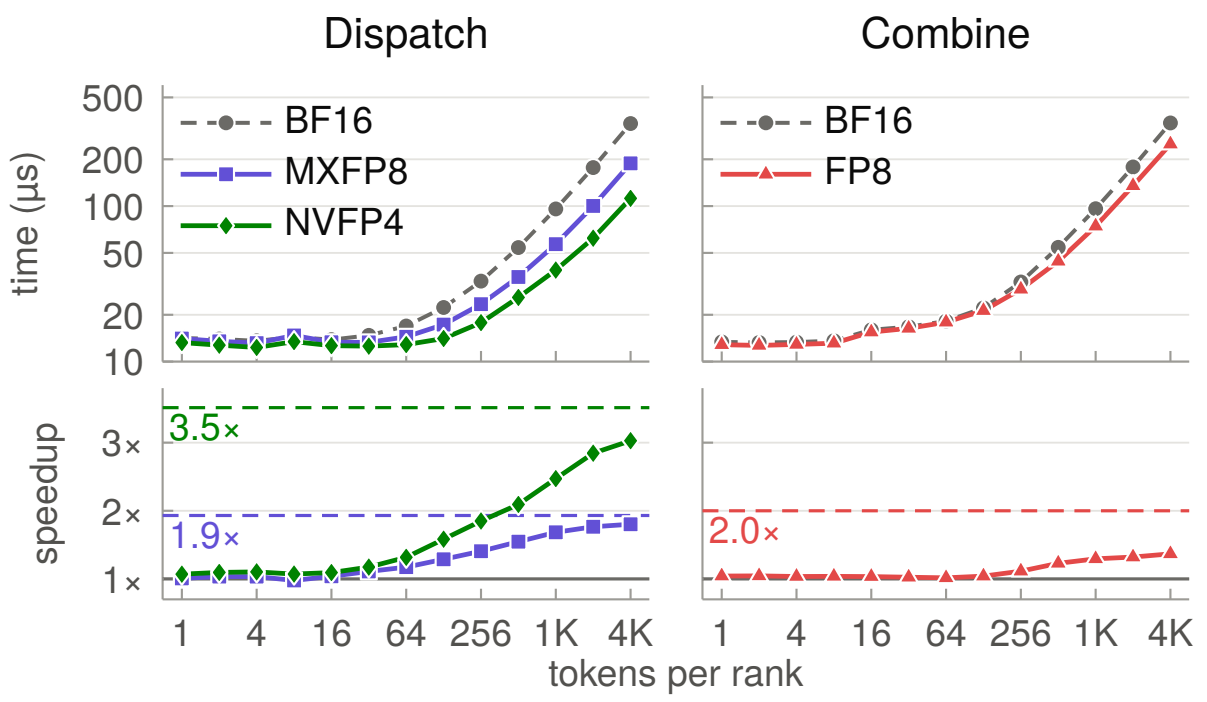
\begin{figure}[t]
  \centering
  \IfFileExists{figures/fig_payload.tex}{\begin{tikzpicture}
\pgfplotsset{
  sBf16/.style={color=cBf16, densely dashed, line width=0.75pt, mark=*, mark size=1.6pt,
    mark options={solid, fill=cBf16, draw=white, line width=0.4pt}},
  sMxfp8/.style={color=cMxfp8, solid, mark=square*, mark size=1.5pt,
    mark options={solid, fill=cMxfp8, draw=white, line width=0.4pt}},
  sNvfp4/.style={color=cNvfp4, solid, mark=diamond*, mark size=2pt,
    mark options={solid, fill=cNvfp4, draw=white, line width=0.4pt}},
  sFp8/.style={color=cFp8, solid, mark=triangle*, mark size=2pt,
    mark options={solid, fill=cFp8, draw=white, line width=0.4pt}},
  payload legend/.style={legend style={at={(0.03,0.97)}, anchor=north west, row sep=-1pt}},
}
\pgfplotstableread{data/series/payload_v41flash_dispatch.dat}\payloaddispatch
\pgfplotstableread{data/series/payload_v41flash_combine.dat}\payloadcombine
\pgfplotstablegetelem{0}{ratio_mxfp8}\of\payloaddispatch \let\ratioMxfp\pgfplotsretval
\pgfplotstablegetelem{0}{ratio_nvfp4}\of\payloaddispatch \let\ratioNvfp\pgfplotsretval
\pgfplotstablegetelem{0}{ratio_fp8}\of\payloadcombine \let\ratioFp\pgfplotsretval
\newcommand{\byteratio}[2]{%
  \draw[#1, densely dashed, line width=0.6pt] (axis cs:0.7,#2) -- (axis cs:5800,#2);
  \node[anchor=north west, font=\plotfont, text=#1, inner sep=1pt] at (axis cs:0.7,#2)
    {\pgfmathprintnumber[fixed, fixed zerofill, precision=1]{#2}\texttimes};}
\begin{groupplot}[
  group style={group size=2 by 2, horizontal sep=10pt, vertical sep=5pt, xticklabels at=edge bottom,
    yticklabels at=edge left, ylabels at=edge left},
  paper, tokens axis, xlabel={}, scale only axis, width=0.39\linewidth,
  xmin=0.7, xmax=5800, xtick={1,4,16,64,256,1024,4096}, xticklabels={1,4,16,64,256,1K,4K},
]
\nextgroupplot[title={Dispatch}, height=0.72in,
  ymode=log, ymin=10, ymax=600, ytick={10,20,50,100,200,500}, yticklabels={10,20,50,100,200,500},
  ylabel={time (\textmu s)}, payload legend]
  \addplot[sBf16] table[x=T, y=bf16] {\payloaddispatch};
  \addlegendentry{BF16}
  \addplot[sMxfp8] table[x=T, y=mxfp8] {\payloaddispatch};
  \addlegendentry{MXFP8}
  \addplot[sNvfp4] table[x=T, y=nvfp4] {\payloaddispatch};
  \addlegendentry{NVFP4}
\nextgroupplot[title={Combine}, height=0.72in,
  ymode=log, ymin=10, ymax=600, ytick={10,20,50,100,200,500}, payload legend]
  \addplot[sBf16] table[x=T, y=bf16] {\payloadcombine};
  \addlegendentry{BF16}
  \addplot[sFp8] table[x=T, y=fp8] {\payloadcombine};
  \addlegendentry{FP8}
\nextgroupplot[height=0.55in, ymin=0.7, ymax=3.8, speedup ticks, ylabel={speedup}]
  \byteratio{cMxfp8}{\ratioMxfp}
  \byteratio{cNvfp4}{\ratioNvfp}
  \addplot[sMxfp8] table[x=T, y=speedup_mxfp8] {\payloaddispatch};
  \addplot[sNvfp4] table[x=T, y=speedup_nvfp4] {\payloaddispatch};
\nextgroupplot[height=0.55in, ymin=0.7, ymax=3.8, speedup ticks, yticklabels={}]
  \byteratio{cFp8}{\ratioFp}
  \addplot[sFp8] table[x=T, y=speedup_fp8] {\payloadcombine};
\end{groupplot}
\node[anchor=north, inner sep=0pt, font=\plotfont, text=cInk2]
  at ($(group c1r2.south)!0.5!(group c2r2.south)+(0,-11pt)$) {tokens per rank};
\end{tikzpicture}}{\figplaceholder{1.6in}{fig\_payload}}
  \caption{Low-precision formats accelerate dispatch and combine at large batches; dashed: the byte ratio, their
  theoretical limit. \sys with \fence, DeepSeek-V4.1-Flash on eight B300 GPUs; top: latency of each phase; bottom:
  speedup over BF16.}
  \label{fig:payload}
\end{figure}

\paragraph{Payload precision.}
\label{sec:eval-precision}
Low-precision payloads pay off at large batches, where \fence is used (Figure~\ref{fig:payload}): at 4096 tokens per
rank, NVFP4 makes dispatch \prelim{3.0$\times$} faster than BF16 for a byte ratio of 3.5$\times$, and MXFP8
\prelim{1.8$\times$} for 1.9$\times$, while at one token all formats take the same time. FP8 combine halves the bytes
but is only \prelim{1.4$\times$} faster, as a kernel must first read and quantize each BF16 partial result; with
per-element E4M3, it shows no statistically significant accuracy loss on downstream tasks.

\subsection{MoE layer}
\label{sec:eval-moe}

\begin{table*}[t]
  \centering
  \footnotesize
  \caption{MoE layer latency (\textmu s) on eight B300 GPUs: layer time and
  dispatch\,/\,MoE compute\,/\,combine stages (MegaMoE: one kernel); fastest layer in bold.}
  \label{tab:moe-layer}
  \setlength{\tabcolsep}{3.5pt}
  \begin{tabular}{@{}r|rc|rc|rc|r@{}}
    \toprule
    & \multicolumn{2}{c|}{\sys\ \fence} & \multicolumn{2}{c|}{\sys\ \sent} & \multicolumn{2}{c|}{\dedg} & MegaMoE \\
    Tokens/rank & MoE layer & stages & MoE layer & stages & MoE layer & stages & MoE layer \\
    \midrule
    \csname @@input\endcsname generated/table_moe_body.tex
    \bottomrule
  \end{tabular}
\end{table*}

Table~\ref{tab:moe-layer} lists the layer and its stages. Although built from separate kernels, our layer is faster
than the fused MegaMoE up to 128 tokens per rank, by up to \prelim{30\%}; from 256 tokens MegaMoE leads, by up to
\prelim{35\%} at 4096, as it hides the communication behind the grouped GEMMs. Our layer is faster than \deepep +
DeepGEMM at every batch, by up to \prelim{2.4$\times$}, mostly through dispatch and combine.

\paragraph{Skewed routing.}
Real routing is imbalanced, and the rank that holds the popular experts limits the layer. We model this with Zipf
routing (Table~\ref{tab:skew}): each token draws $\topk$ distinct experts with probability proportional to
$i^{-\alpha}$, where $i$ is the expert's position in a random order ($\alpha=0$: uniform). At one token per rank, the
grouped GEMMs are bound by loading expert weights, so skew, which activates fewer experts, can even make the layers
faster. At 1024 tokens per rank, the busiest rank computes up to \prelim{1.9} times as many token--expert routing pairs as a
rank under perfect routing (load imbalance in Table~\ref{tab:skew}), and its grouped GEMMs slow all three layers by
\prelim{50--70\%}. Our communication grows less: a token reaches a target rank once, however many of the rank's
experts it selects, so the bytes a rank receives follow its tokens, at most $\nranks/\min(\topk,\nranks)$ times as many
as under perfect routing, rather than its token--expert routing pairs. At $\alpha=1.25$, dispatch and combine of the
busiest rank take \prelim{160} and \prelim{135}\us, against \prelim{267} and
\prelim{400}\us with \deepep + DeepGEMM, which copies each token once per selected local expert.

\begin{table*}[t]
  \centering
  \footnotesize
  \caption{MoE layer under Zipf routing with exponent $\alpha$ ($\alpha=0$: uniform) on eight B300 GPUs, latency in
  \textmu s. Imba.\ (load imbalance) is the ratio of the token--expert routing pairs that the busiest rank computes to
  those that each rank computes under perfect routing; 1 means balanced. Communication is the dispatch and combine of
  the busiest rank. Ours: \sys with the faster protocol + Prims-TS; DE+DG: \deepep + DeepGEMM.}
  \label{tab:skew}
  \setlength{\tabcolsep}{2.5pt}
  \begin{tabular}{@{}rr|rrr|rrr|rrr|rrr@{}}
    \toprule
    & & \multicolumn{6}{c|}{DeepSeek-V4.1-Flash} & \multicolumn{6}{c}{Kimi-K3} \\
    & & & \multicolumn{2}{c|}{Communication} & \multicolumn{3}{c|}{MoE layer} & & \multicolumn{2}{c|}{Communication} &
      \multicolumn{3}{c}{MoE layer} \\
    Tokens/rank & $\alpha$ & \loadimb & Ours & \deepep & Ours & \dedg & MegaMoE & \loadimb & Ours & \deepep & Ours & \dedg &
      MegaMoE \\
    \midrule
    \csname @@input\endcsname generated/table_skew_body.tex
    \bottomrule
  \end{tabular}
\end{table*}

\subsection{Large EP scaling}
\label{sec:eval-scaling}

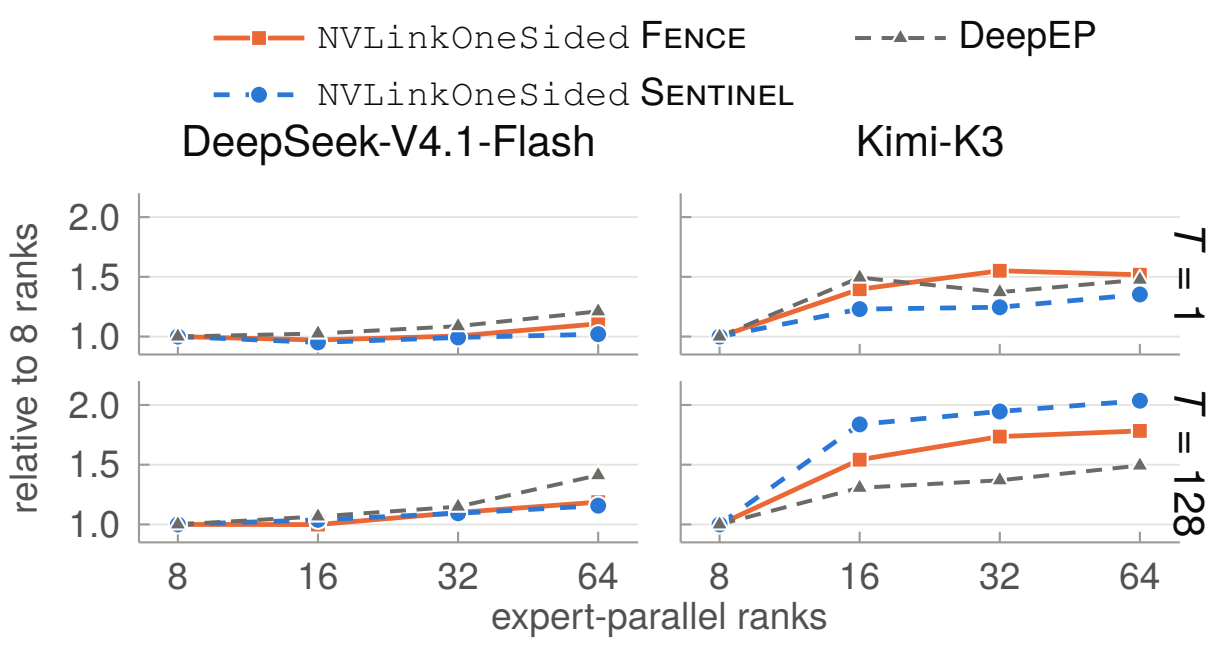
\begin{figure}[t]
  \centering
  \begin{tikzpicture}
\pgfplotsset{sSentinelDashed/.style={sSentinel, dashed}}
\newcommand{\scalingplot}[3]{%
  \pgfplotstableread{#3}\scalingtable
  \pgfplotstablegetelem{0}{#2}\of\scalingtable
  \edef\scalingnext{\noexpand\addplot[#1] table[x=EP, y expr=\noexpand\thisrow{#2}/\pgfplotsretval] {#3};}%
  \scalingnext}
\begin{groupplot}[
  group style={group size=2 by 2, horizontal sep=8pt, vertical sep=5pt, xticklabels at=edge bottom,
    yticklabels at=edge left},
  paper, scale only axis, width=0.4\linewidth, height=0.42in,
  xmode=log, xmin=6.6, xmax=78, xtick={8,16,32,64}, xticklabels={8,16,32,64},
  ymin=0.85, ymax=2.2, ytick={1,1.5,2}, yticklabel style={/pgf/number format/.cd, fixed, fixed zerofill, precision=1},
]
\nextgroupplot[title={DeepSeek-V4.1-Flash}, legend to name=scalinglegend, legend columns=2,
  transpose legend]
  \scalingplot{sFence}{fence}{data/series/scaling_v41flash_t1.dat}
  \addlegendentry{\sys\ \fence}
  \scalingplot{sSentinelDashed}{sentinel}{data/series/scaling_v41flash_t1.dat}
  \addlegendentry{\sys\ \sent}
  \scalingplot{sDeepEp}{deepep}{data/series/scaling_v41flash_t1.dat}
  \addlegendentry{\deepep}
\nextgroupplot[title={Kimi-K3}]
  \scalingplot{sFence}{fence}{data/series/scaling_kimik3_t1.dat}
  \scalingplot{sSentinelDashed}{sentinel}{data/series/scaling_kimik3_t1.dat}
  \scalingplot{sDeepEp}{deepep}{data/series/scaling_kimik3_t1.dat}
\nextgroupplot[]
  \scalingplot{sFence}{fence}{data/series/scaling_v41flash_t128.dat}
  \scalingplot{sSentinelDashed}{sentinel}{data/series/scaling_v41flash_t128.dat}
  \scalingplot{sDeepEp}{deepep}{data/series/scaling_v41flash_t128.dat}
\nextgroupplot[]
  \scalingplot{sFence}{fence}{data/series/scaling_kimik3_t128.dat}
  \scalingplot{sSentinelDashed}{sentinel}{data/series/scaling_kimik3_t128.dat}
  \scalingplot{sDeepEp}{deepep}{data/series/scaling_kimik3_t128.dat}
\end{groupplot}
\node[anchor=south, inner sep=0pt] at ($(group c1r1.north)!0.5!(group c2r1.north)+(0,13pt)$)
  {\pgfplotslegendfromname{scalinglegend}};
\node[rotate=-90, anchor=north, font=\plottitlefont, text=cInk, inner sep=0pt] at ($(group c2r1.east)+(5pt,0)$)
  {$\ntok=1$};
\node[rotate=-90, anchor=north, font=\plottitlefont, text=cInk, inner sep=0pt] at ($(group c2r2.east)+(5pt,0)$)
  {$\ntok=128$};
\node[rotate=90, anchor=south, font=\plotfont, text=cInk2, inner sep=0pt]
  at ($(group c1r1.south west)!0.5!(group c1r2.north west)+(-19pt,0)$) {relative to 8 ranks};
\node[anchor=north, inner sep=0pt, font=\plotfont, text=cInk2]
  at ($(group c1r2.south)!0.5!(group c2r2.south)+(0,-11pt)$) {expert-parallel ranks};
\end{tikzpicture}
  \caption{At one token per rank, \sent's round trip grows least from 8 to 64 GPUs. All-to-all
  round trip on NVL72 relative to the same implementation on 8 GPUs, at a fixed number of tokens per rank.}
  \label{fig:scaling}
\end{figure}

A larger EP group gives each GPU fewer experts, which frees memory and, at small batches, where the grouped GEMMs are
bound by loading weights, reduces the weights each GPU loads. We scale EP from 8 to 64 GPUs on one NVL72 rack at a
fixed number of tokens per rank.

The all-to-all slows only slightly (Figure~\ref{fig:scaling}). At one token per rank on DeepSeek-V4.1-Flash, \sent's
round trip stays at \prelim{16.6--17.8}\us, as it has no per-peer handshake, while \fence's grows from \prelim{43.0} to
\prelim{47.6}\us, as it publishes and awaits one flag per peer. Kimi-K3 scales worse, as from 16 GPUs on its tokens ($\topk=16$) reach 16
ranks instead of eight.

The MoE layer gains at intermediate batches (Table~\ref{tab:gb200-body}): on 64 GPUs, a rank holds six experts of
DeepSeek-V4.1-Flash instead of 48, and at 128 tokens per rank our layer takes \prelim{103} instead of \prelim{182}\us.
At one token per rank, a rank activates about $\topk$ experts at any EP size, and at large batches it processes
$\ntok\topk$ token--expert routing pairs at any EP size, so large EP does not reduce its work in either case. Our layer stays
\prelim{23--26\%} below MegaMoE at one token per rank, while MegaMoE gains more at intermediate batches and leads from
\prelim{32} tokens per rank on 64 GPUs instead of from \prelim{128} on eight.

\begin{table}[t]
  \centering
  \footnotesize
  \caption{MoE layer latency (\textmu s) of DeepSeek-V4.1-Flash on one GB200 NVL72 rack, by EP size; Ours: the faster
  protocol. Fastest layer in bold.}
  \label{tab:gb200-body}
  \setlength{\tabcolsep}{4pt}
  \begin{tabular}{@{}rlrrrr@{}}
    \toprule
    & & \multicolumn{4}{c}{EP size (GPUs)} \\
    \cmidrule(l){3-6}
    $\ntok$ & & 8 & 16 & 32 & 64 \\
    \midrule
    \csname @@input\endcsname generated/table_gb200_body.tex
    \bottomrule
  \end{tabular}
\end{table}

\section{Discussion}
\label{sec:discussion}

\paragraph{Advantages and limitations.}
\sys performs well across the board: its all-to-all is faster than \deepep at every batch, and its layer, built from
separate kernels, is faster than the fused MegaMoE up to 128 tokens per rank on eight GPUs (\S\ref{sec:eval}). Keeping the
all-to-all separate from MoE compute makes it adaptable: it carries any quantization format and pairs with various MoE
compute backends behind one interface, and it leaves room for features of the communication alone, such as fault tolerance.
Its main limitation is that the EP group must fit in one NVLink domain: both protocols rely on direct loads and stores
to peer memory and on 32-bit words that arrive whole, which RDMA does not offer (\S\ref{sec:bg-nvlink}). Its dispatch
workspace holds one region per rank of the group, so its memory grows with the EP size.

\paragraph{MegaMoE.}
DeepGEMM MegaMoE fuses dispatch, both grouped GEMMs and combine into one persistent kernel that
exchanges token indices first and then pulls the tokens into expert-major blocks sized for its GEMM tiles. Its
strength is overlap: within one kernel, NVLink transfers proceed during Tensor Core work, which separate kernels
cannot do, so MegaMoE leads from 256 tokens per rank (\S\ref{sec:eval-moe}). Its weaknesses are the metadata exchange
ahead of the data and its coupling:
layout, transfer granularity and formats follow the GEMM's tiles, so a new grouped GEMM or precision needs a new
implementation. In future work, we plan to carry the one-sided design into such a fused operator, pushing tokens
without metadata into a rank-major layout and detecting their arrival from the data, so that communication also
overlaps with computation at large batches.

\section{Related Work}
\label{sec:related}

\paragraph{MoE systems.}
Expert parallelism was introduced with GShard~\cite{lepikhin2021gshard} and Switch Transformer~\cite{fedus2022switch}.
DeepSpeed-MoE~\cite{rajbhandari2022deepspeedmoe} co-designs the model with an inference system that combines expert,
tensor and data parallelism with hierarchical all-to-all; FasterMoE~\cite{he2022fastermoe} counters load imbalance by
replicating popular experts and overlaps communication through fine-grained scheduling; Tutel~\cite{hwang2023tutel}
switches parallelism and pipelining at runtime; and MegaBlocks~\cite{gale2023megablocks} computes experts as
block-sparse operations, so that no token is dropped. MegaScale-Infer~\cite{zhu2025megascaleinfer} places attention
and experts on separate GPUs, possibly of different types, hides the traffic between them with a ping-pong pipeline
over micro-batches, and carries it with an RDMA library from the attention nodes to the expert nodes.

\paragraph{Expert-parallel communication.}
\deepep V1~\cite{deepep} has two kernel families. Its high-throughput kernels, for training and prefill, first
exchange token counts and then forward the tokens between the NVLink and RDMA domains; its low-latency kernels, for
decoding, send each token over RDMA, later also over NVLink, into per-expert receive buffers without a separate
count exchange, and are compatible with CUDA graphs. \deepep V2 rebuilds both on NCCL's GPU-Initiated Networking
(GIN)~\cite{hamidouche2025gin} behind one buffer interface, with an expanded per-expert output layout for grouped GEMMs.
Perplexity's kernels~\cite{pplxkernels,pplx-blog} split sending and receiving into separate kernels and deliver
per-expert outputs, through NVSHMEM across nodes and direct stores within a node. NCCL EP~\cite{ncclep} offers both
modes on NCCL's device API; its low-latency mode can also return the tokens in a rank-major layout. UB-X~\cite{ncclubx},
an experimental NVLink library in NCCL, computes dispatch offsets from a routing matrix that every rank holds and
provides a combine that pushes with sentinel synchronization.

\paragraph{Collective protocols.}
NCCL's LL and LL128 protocols~\cite{nccl,demystifying-nccl} and MSCCL++'s LL packets~\cite{mscclpp} pair data with
flag words so that receivers detect arrival without fences, at the cost of transferring the flags; NVSHMEM's
put-with-signal instead pairs each transfer with a separate signal word~\cite{nvshmem}. \citet{shen2026microsecond}
measure memory barriers at over a microsecond each on GB200 and build barrier-free AllReduce from LL and sentinel
synchronization on NCCL's device API. In dispatch and combine, sentinel synchronization meets two new problems:
quantized payloads may take the sentinel's bit pattern, and the slots that receive tokens on a rank depend on the routing
(\S\ref{sec:sync-sentinel}). In-band completion over one-sided writes also appears in RDMA systems, which validate
remote objects or messages by per-cache-line versions or by the arrival of their last bytes~\cite{farm,herd}, and HPC
separates completion from data transfer through MPI-3 epochs~\cite{mpi3} and Notified Access~\cite{belli2015notified}.

\paragraph{Overlap and fusion.}
Comet~\cite{zhang2025comet} overlaps expert computation with communication at tile granularity, and
TileLink~\cite{zheng2025tilelink} generates overlapped kernels from tile-level primitives.
FlashDMoE~\cite{aimuyo2025flashdmoe} and DeepGEMM MegaMoE~\cite{deepgemm} fuse dispatch, expert computation and
combine into one persistent kernel, and FarSkip-Collective~\cite{farskip2026} changes the model so that blocking
collectives overlap with computation. These reduce how much of the communication is exposed; \sys reduces the cost of
the communication itself, which matters most in decoding, where there is little computation to hide it behind.

\section{Conclusion}
\label{sec:conclusion}

\sys implements dispatch and combine as one-sided communication within an NVLink domain: each source rank pushes its
tokens into rank-major workspaces using only its own routing information, MoE compute reads them through slot indices,
and two protocols, \sent for small batches and \fence for large ones, synchronize the exchange. On Blackwell GPUs, \sys
is faster than \deepep at every batch, and the resulting MoE layer, built from separate kernels, achieves better performance than DeepGEMM MegaMoE at decoding batches. \sys is available in FlashInfer\ifdefined\arxivbuild\footnote{\url{https://docs.flashinfer.ai/generated/flashinfer.moe_ep.NVLinkOneSidedAlltoAll.html}}\fi{} and is integrated into vLLM~\cite{kwon2023vllm},
SGLang~\cite{zheng2024sglang} and TensorRT-LLM~\cite{trtllm}.

\section*{Acknowledgements}

We thank Itamar Rosenfeld Rauch, Chanan Maoz, Samuel Nordmann, Nikita Korobov, Yicheng Qiang, Ke Wen, Xiaofan Li,
Jack Yang, Hao Lu and Ao Tang for their help and discussions.

\bibliography{refs}
\bibliographystyle{mlsys2025}

\section*{AI Usage Disclosure}

AI coding assistants were used to help draft and edit the text and to write the plotting and benchmark scripts. All technical content, the design, the experiments and the reported
numbers were produced or verified by the authors. No experimental data was generated by AI.

\clearpage
\appendix
\numberwithin{table}{section}

\section{Latency tables}
\label{app:tables}

Tables~\ref{tab:full-v41flash} and \ref{tab:full-kimik3} list dispatch, combine and round trip of the all-to-all
(\S\ref{sec:eval-latency}) for \sent, \fence and \deepep from 1 to 4096 tokens per rank. Tables~\ref{tab:moe-v41flash}
and \ref{tab:moe-kimik3} list the MoE layer (\S\ref{sec:eval-moe}) with each protocol, MegaMoE and \deepep +
DeepGEMM; Table~\ref{tab:precision} gives the payload formats (Figure~\ref{fig:payload}), and Table~\ref{tab:zipf}
the layer under Zipf routing. On one GB200 NVL72 rack (\S\ref{sec:eval-scaling}), Table~\ref{tab:gb200-scaling} lists
the all-to-all and Table~\ref{tab:gb200-layer} the layer with perfect routing.

\newcommand{\atatable}[3]{%
\begin{table*}[h]
  \centering
  \footnotesize
  \caption{All-to-all latency (\textmu s), #1, 8$\times$B300: dispatch / combine / round trip.}
  \label{#2}
  \setlength{\tabcolsep}{4pt}
  \begin{tabular}{@{}r|rrr|rrr|rrr@{}}
    \toprule
    & \multicolumn{3}{c|}{\sent} & \multicolumn{3}{c|}{\fence} & \multicolumn{3}{c}{\deepep} \\
    Tokens/rank & disp. & comb. & total & disp. & comb. & total & disp. & comb. & total \\
    \midrule
    \csname @@input\endcsname #3.tex
    \bottomrule
  \end{tabular}
\end{table*}}
\atatable{DeepSeek-V4.1-Flash}{tab:full-v41flash}{generated/table_b300_v41flash}
\atatable{Kimi-K3}{tab:full-kimik3}{generated/table_b300_kimik3}

\newcommand{\moetable}[3]{%
\begin{table*}[h]
  \centering
  \footnotesize
  \caption{MoE layer latency (\textmu s), #1, 8$\times$B300, MXFP4 weights and MXFP8 activations: layer time and
  dispatch / MoE compute / combine, each stage from its first kernel's start to its last kernel's end.}
  \label{#2}
  \setlength{\tabcolsep}{3.5pt}
  \begin{tabular}{@{}r|rc|rc|r|rc@{}}
    \toprule
    & \multicolumn{2}{c|}{\sys\ \fence} & \multicolumn{2}{c|}{\sys\ \sent} & MegaMoE &
    \multicolumn{2}{c}{\dedg} \\
    Tokens/rank & MoE layer & stages & MoE layer & stages & MoE layer & MoE layer & stages \\
    \midrule
    \csname @@input\endcsname #3.tex
    \bottomrule
  \end{tabular}
\end{table*}}
\moetable{DeepSeek-V4.1-Flash}{tab:moe-v41flash}{generated/table_moe_v41flash}
\moetable{Kimi-K3}{tab:moe-kimik3}{generated/table_moe_kimik3}

\begin{table}[h]
  \centering
  \footnotesize
  \caption{\fence with other payload formats on eight B300 GPUs: bytes per token and target rank, and round trip
  (\textmu s). The last two rows of each model include quantization to MXFP8, on the source rank or, after a BF16
  dispatch, on the target rank.}
  \label{tab:precision}
  \setlength{\tabcolsep}{2.5pt}
  \begin{tabular}{@{}llrrrrr@{}}
    \toprule
    Dispatch & Combine & Bytes & \multicolumn{4}{c}{Tokens per rank} \\
    \cmidrule(l){4-7}
    & & disp.\,/\,comb. & 1 & 128 & 1024 & 4096 \\
    \midrule
    \csname @@input\endcsname generated/table_precision.tex
    \bottomrule
  \end{tabular}
\end{table}

\begin{table*}[h]
  \centering
  \footnotesize
  \caption{MoE layer under Zipf routing with exponent $\alpha$ on eight B300 GPUs: layer time (\textmu s) of Ours, DE+DG
  and MegaMoE. Ratio is ours relative to MegaMoE, and Imba.\ is the load imbalance of Table~\ref{tab:skew}. Ours: \sys
  with the faster protocol + Prims-TS ($^\dagger$: \fence); DE+DG: \deepep + DeepGEMM.}
  \label{tab:zipf}
  \setlength{\tabcolsep}{4pt}
  \begin{tabular}{@{}lrrrrrrrrrrr@{}}
    \toprule
    & & \multicolumn{5}{c}{DeepSeek-V4.1-Flash} & \multicolumn{5}{c}{Kimi-K3} \\
    \cmidrule(lr){3-7}\cmidrule(l){8-12}
    Routing & $\ntok$ & Ours & \dedg & MegaMoE & Ratio & \loadimb & Ours & \dedg & MegaMoE & Ratio & \loadimb \\
    \midrule
    \csname @@input\endcsname generated/table_zipf.tex
    \bottomrule
  \end{tabular}
\end{table*}

\begin{table*}[h]
  \centering
  \footnotesize
  \caption{All-to-all round trip (\textmu s, dispatch + combine) on one GB200 NVL72 rack with perfect routing, by
  expert-parallel ranks (on ranks\,/\,4 compute trays) and tokens per rank.}
  \label{tab:gb200-scaling}
  \setlength{\tabcolsep}{6pt}
  \begin{tabular}{@{}rrrrrrrr@{}}
    \toprule
    & & \multicolumn{3}{c}{DeepSeek-V4.1-Flash} & \multicolumn{3}{c}{Kimi-K3} \\
    \cmidrule(lr){3-5}\cmidrule(l){6-8}
    Ranks & $\ntok$ & \fence & \sent & \deepep & \fence & \sent & \deepep \\
    \midrule
    \csname @@input\endcsname generated/table_gb200_scaling.tex
    \bottomrule
  \end{tabular}
\end{table*}

\begin{table*}[h]
  \centering
  \footnotesize
  \caption{MoE layer latency (\textmu s) on one GB200 NVL72 rack with perfect routing, by expert-parallel ranks and
  tokens per rank: \sys (the faster protocol; $^\dagger$: \fence), \deepep + DeepGEMM and MegaMoE. --: not tested,
  as the Prims-TS grouped GEMMs need at least $\topk$ experts per rank (Kimi-K3 has 14 on 64 GPUs).}
  \label{tab:gb200-layer}
  \setlength{\tabcolsep}{6pt}
  \begin{tabular}{@{}rrrrrrrr@{}}
    \toprule
    & & \multicolumn{3}{c}{DeepSeek-V4.1-Flash} & \multicolumn{3}{c}{Kimi-K3} \\
    \cmidrule(lr){3-5}\cmidrule(l){6-8}
    Ranks & $\ntok$ & Ours & \dedg & MegaMoE & Ours & \dedg & MegaMoE \\
    \midrule
    \csname @@input\endcsname generated/table_gb200_layer.tex
    \bottomrule
  \end{tabular}
\end{table*}

\section{Measurement details}
\label{app:method}

The B300 node runs CUDA 13.5, and the tokens are quantized to MXFP8 before the timed span. Ten warm-up and 100 timed
iterations run back to back in one CUDA graph, with L2 flushed before each; we report the span from the first to the
last kernel, averaged over iterations and ranks. All-to-all outputs match a reference bitwise, and every layer matches
one to a relative L2 error of at most \prelim{$5.7\times10^{-3}$}.
We time phases from CUPTI kernel records in back-to-back CUDA graph replays, with no barrier or event between phases:
a barrier before every phase hides the waiting a protocol does in steady state, and events in the graph keep
Programmatic Dependent Launch (PDL) from overlapping the kernels around them. The round trip of the all-to-all is the sum
of dispatch and combine. Each stage of the MoE layer (Table~\ref{tab:moe-layer}) spans its first kernel's start to its
last kernel's end; PDL lets stages overlap slightly, so they add up to more than the layer.
The breakdown in Figure~\ref{fig:sync} comes from IKET timestamps that every warp records when it enters and leaves
each step, traced back from the block that finishes each phase last and averaged over ranks and iterations;
instrumentation changes the round trip by at most \prelim{6\%}.

\end{document}